\documentclass[aps,prx,superscriptaddress,twocolumn,amsmath,longbibliography,showkeys]{revtex4-2}

\usepackage{amsmath}
\usepackage{amssymb}
\usepackage{graphicx}
\usepackage{dcolumn}
\usepackage{bm}
\usepackage{hyperref}

\usepackage[usenames,dvipsnames]{xcolor}
\newcommand{\GL}[1]{{\color{black}#1}}

\begin{document}


\title{Selective-Area Epitaxy of Bulk-Insulating (Bi$_\text{x}$Sb$_\text{1-x}$)$_2$Te$_3$ Films and Nanowires by Molecular Beam Epitaxy}

\author{Gertjan Lippertz}
\affiliation{Physics Institute II, University of Cologne, D-50937 K{\"o}ln, Germany}

\author{Oliver Breunig}
\affiliation{Physics Institute II, University of Cologne, D-50937 K{\"o}ln, Germany}

\author{Rafael Fister}
\affiliation{Physics Institute II, University of Cologne, D-50937 K{\"o}ln, Germany}

\author{Anjana Uday}
\affiliation{Physics Institute II, University of Cologne, D-50937 K{\"o}ln, Germany}

\author{Andrea Bliesener}
\affiliation{Physics Institute II, University of Cologne, D-50937 K{\"o}ln, Germany}

\author{\GL{Jens Brede}}
\affiliation{Physics Institute II, University of Cologne, D-50937 K{\"o}ln, Germany}

\author{Alexey Taskin}
\affiliation{Physics Institute II, University of Cologne, D-50937 K{\"o}ln, Germany}

\author{Yoichi Ando} \email{ando@ph2.uni-koeln.de}
\affiliation{Physics Institute II, University of Cologne, D-50937 K{\"o}ln, Germany}

\begin{abstract}
The selective-area epitaxy (SAE) is a useful technique to grow epitaxial films with a desired shape on a pre-patterned substrate. Although SAE of patterned topological-insulator (TI) thin films has been performed in the past, there has been no report of SAE-grown TI structures that are bulk-insulating. Here we report the successful growth of Hall-bars and nanowires of bulk-insulating TIs using the SAE technique. Their transport properties show that the quality of the selectively-grown structures is comparable to that of bulk-insulating TI films grown on pristine substrates. In SAE-grown TI nanowires, we were able to observe Aharonov-Bohm-like magnetoresistance oscillations that are characteristic of the quantum-confined topological surface states. The availability of bulk-insulating TI nanostructures via the SAE technique opens the possibility to fabricate intricate topological devices in a scalable manner.
 \end{abstract}

\keywords{topological insulator, nanowire, selective area epitaxy, Aharonov-Bohm oscillations, electrostatic gating}

\maketitle

\newpage

\section{\label{sec:intro}Introduction}

Topological insulators (TIs) have a high application potential when used in devices for spintronics or topological quantum computing \cite{Breunig2022}. Such device applications often require patterned TI films. In this context, the selective area epitaxy (SAE) of TIs has been prolific in recent years, allowing for the realization of high-precision and scalable nanostructures. Examples include TI nanoribbons \cite{Rosenbach2020,Rosenbach2022,Koelzer2021}, nanorings \cite{Behner2023}, in-situ fabricated Josephson junctions \cite{Schueffelgen2019,Koelzer2023}, and  TI-based qubit circuits \cite{Schmitt2022}. The SAE technique relies on a prepatterned growth template consisting of two material systems. One of these materials is single-crystalline and lattice-matched to the epilayer to be selectively grown, whereas the second material is often amorphous and does not allow the heterogeneous nucleation of the epilayer to take place. 

The TI material (Bi$_\text{x}$Sb$_\text{1-x}$)$_2$Te$_3$ is of particular interest for applications, because it can be made bulk-insulating with a careful tuning of the growth condition and the Sb/Bi ratio \cite{Kong2011,Zhang2011}.
For the SAE growth of (Bi$_\text{x}$Sb$_\text{1-x}$)$_2$Te$_3$, a natural choice would be to combine Si(111) substrate (which is known to support the epitaxial growth of this material) with its native oxide SiO$_2$ acting as the growth mask. While SAE is indeed possible on Si(111)-SiO$_2$ templates \cite{Lanius2018}, Jalil \textit{et al.} showed that a better dimensional control can be achieved at the nanoscale when amorphous Si$_3$N$_4$ is used as the growth mask instead \cite{Jalil2023}. However, for the epitaxial growth on the Si(111) surface a few challenges remain: (i) X-ray diffraction has shown that the crystal quality of (Bi$_\text{x}$Sb$_\text{1-x}$)$_2$Te$_3$ thin films grown on Si(111) is inferior when compared to other growth substrates \cite{Park2012,Richardson2017}. (ii) Atom-probe tomography has revealed Sb accumulation at the film-substrate interface, which may give rise to a parasitic two-dimensional electron gas, since Sb is a well-known $n$-type dopant in Si \cite{Lanius2016}.

It is useful to emphasize that bulk-insulation is important for TI materials in order to access the topological phenomena, which often requires the chemical potential to be tuned close to the Dirac point of the topological surface state. For example, in the case of a TI proximitized by a superconductor to realize topological superconductivity, the Majorana zero modes appearing in such a system become more stable as the chemical potential gets closer to the Dirac point \cite{Fu2008, Legg2021}. While bulk-insulation can in principle be achieved in TI thin films by taking (Bi$_\text{x}$Sb$_\text{1-x}$)$_2$Te$_3$ as the material platform, it actually requires a fine tuning of the conditions of the molecular beam epitaxy (MBE) \cite{Kong2011, Zhang2011, Yang2015}. Since a successful SAE requires another fine tuning of the MBE conditions, it is very difficult to find the right condition to achieve both SAE and bulk-insulation. As a result, there has been no report of bulk-insulating TI structure grown via the SAE technique \cite{Rosenbach2020,Rosenbach2022,Koelzer2021,Behner2023,Schueffelgen2019,Koelzer2023,Schmitt2022,Lanius2016,Jalil2023}. 

In this work, we succeeded for the first time in the SAE of bulk-insulating (Bi$_\text{x}$Sb$_\text{1-x}$)$_2$Te$_3$ structures (Hall-bar and nanowire) on patterned Al$_2$O$_3$(0001)-Si$_3$N$_4$ substrates. The advantage of the epitaxial growth on sapphire is that it does not suffer from the issues mentioned above for Si substrates. Moreover, the SAE can be achieved in a simple bilayer structure, whereas Jalil \textit{et al.} had to rely on a thin SiO$_2$ buffer layer between the Si(111) surface and Si$_3$N$_4$ mask to mitigate the tensile strain exerted by the Si$_3$N$_4$ layer on the Si interface \cite{Jalil2023}. Strain is not an issue for sapphire substrates owing to its large Young's modulus. By optimizing the MBE growth condition, we achieved a reasonably high carrier mobility of $\sim$1000 cm$^2$/Vs comparable to plain films grown on pristine substrates \cite{Yang2015}. This is nontrivial, because the growth substrate for SAE is first completely covered with Si$_3$N$_4$ and then trenches are patterned in a nanofabrication process. In our experiment, we found that the realization of bulk-insulation for nano-sized structures requires re-optimisation of the growth conditions that turned out to depend on the feature size. In the bulk-insulating TI nanowires obtained via SAE, our transport experiments found Aharonov-Bohm (AB)-like magnetoresistance oscillations in magnetic fields applied parallel to the nanowire, that are characteristic of the quasi-one-dimensional transport through size-quantized topological surface states \cite{Peng2010, Xiu2011, Cho2015, Roessler2023}, confirming their topological nature.

\begin{figure}
\centering
\includegraphics[width=.45\textwidth]{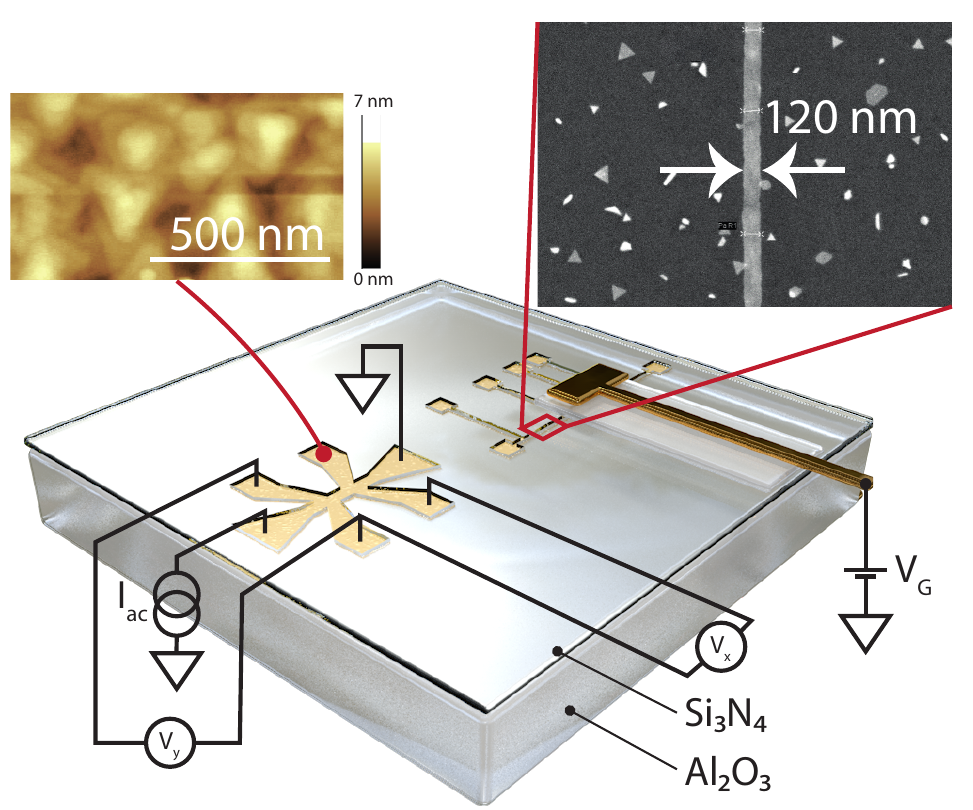}
\vspace*{0mm}\caption{Schematic of a substrate used for SAE, consisting of a sapphire (Al$_2$O$_3$) substrate and a patterned silicon nitride (Si$_3$N$_4$) mask with a Hall-bar design (front) and a multi-terminal nanowire (back). Dimensions are not to scale. The MBE film (yellow) grows only on the exposed sapphire region within the Si$_3$N$_4$ mask openings. A top-gate voltage $V_\mathrm{G}$ is applied to gold electrodes fabricated on top of an ALD-deposited Al$_2$O$_3$ dielectric (exemplary shown here only for parts of the nanowire). For transport measurements an ac current is applied to the Hall-bar between the current electrodes. The longitudinal voltage $V_\mathrm{x}$ is acquired together with the Hall voltage $V_\mathrm{y}$ using lockin amplifiers. Upper left inset: AFM image of the MBE film selectively grown in one of the larger pad areas. Upper right inset: SEM image of a 120-nm-wide nanowire. Only few isolated crystallites are found on the Si$_3$N$_4$ mask (black).}
\label{fig:1}
\end{figure}

\section{\label{sec:exp}Results and Discussion}

By applying a two-step temperature profile to the prepatterned substrate during the MBE growth (details provided in Methods), the SAE growth of (Bi$_\text{x}$Sb$_\text{1-x}$)$_2$Te$_3$ Hall-bars and nanowires is achieved. Figure \ref{fig:1} shows a schematic of the Al$_2$O$_3$ substrate with Si$_3$N$_4$ mask, as well as an atomic-force-microscopy (AFM) and scanning-electron-microscopy (SEM) images of the selectively-grown (Bi$_\text{x}$Sb$_\text{1-x}$)$_2$Te$_3$ structures. The film in the Hall-bar is continuous with triangular terraces, indicating a good epitaxy. This growth of continuous (Bi$_\text{x}$Sb$_\text{1-x}$)$_2$Te$_3$ is maintained even when the lateral dimensions of the exposed areas in the growth mask are constrained into nanowires as seen from the SEM image. Due to a high degree of selectivity, only a few isolated crystallites are found on the Si$_3$N$_4$ mask.

\begin{figure*}
\centering
\includegraphics[width=.75\textwidth]{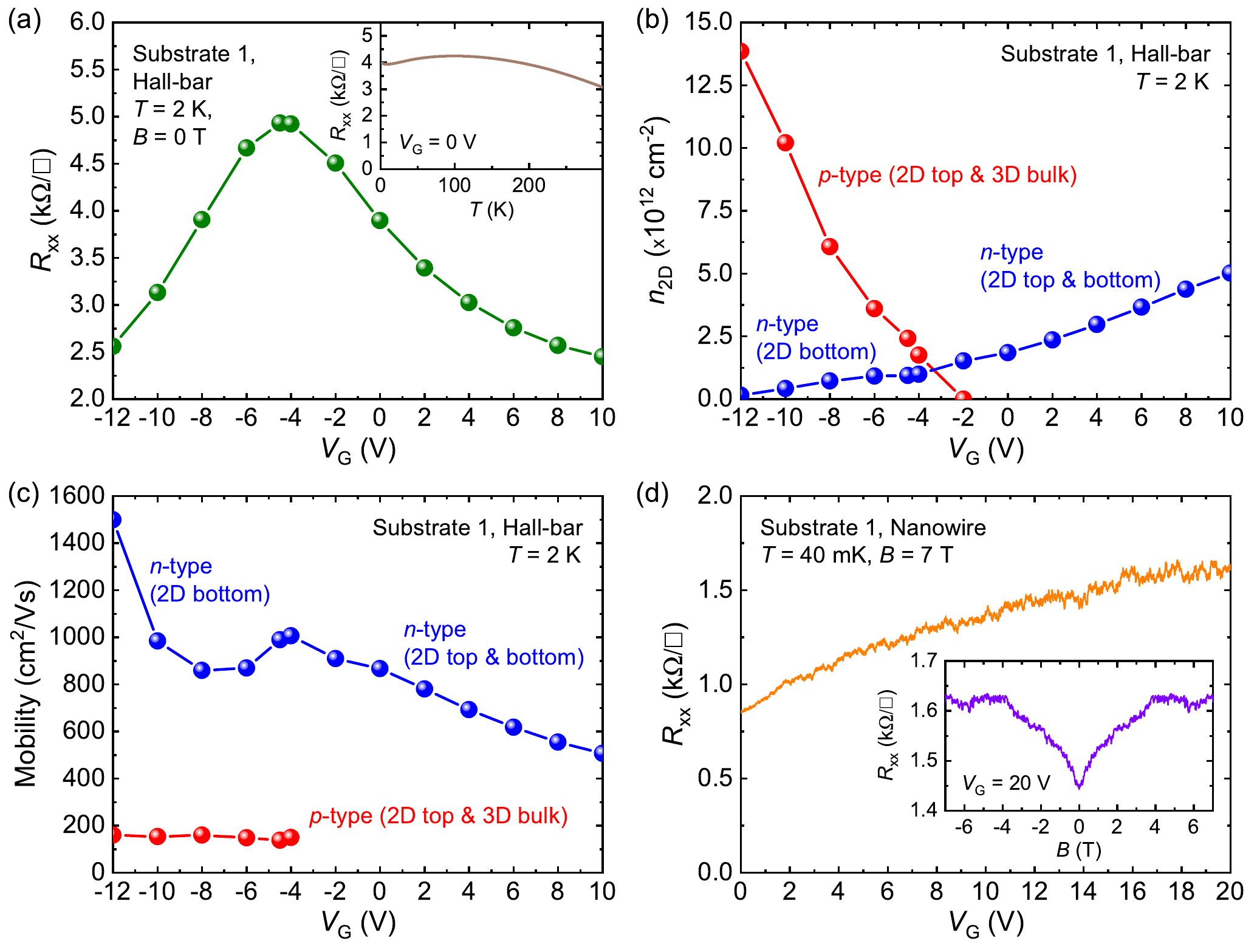}
\vspace*{0mm}\caption{Transport characterization of the (Bi$_\text{x}$Sb$_\text{1-x}$)$_2$Te$_3$ devices grown on the SAE substrate 1. (a) $V_\text{G}$-dependence of $R_\text{xx}$ of the Hall-bar at $2\,$K. Inset: Temperature dependence of $R_\text{xx}$ at $V_\text{G}=0\,$V, showing a typical bulk-insulating behavior. (b,c) $n_\text{2D}$ and mobility obtained for the Hall-bar as a function of $V_\text{G}$ at $2\,$K from the two-band analysis of the $R_{yx}(B)$ data shown in the SI. (d) $V_\text{G}$-dependence of $R_\text{xx}$ (taken at $40\,$mK in the parallel magnetic field of $7\,$T) of a 160-nm-wide nanowire on the same substrate, showing a $p$-type behavior within the experimental gating range. Inset: magnetoresistance of the same nanowire in parallel magnetic fields up to $7\,$T measured at $V_\text{G}=20\,$V and $T=40\,$mK.}
\label{fig:2}
\end{figure*}

To characterize the electronic properties of the (Bi$_\text{x}$Sb$_\text{1-x}$)$_2$Te$_3$ film selectively-grown in the Hall-bar structure, the longitudinal sheet resistance $R_\text{xx}$ and the Hall resistance $R_\text{yx}$ were measured as a function of temperature $T$, perpendicular magnetic field $B$, and top-gate voltage $V_{\rm G}$. \GL{We follow the convention to express the sheet resistance in the unit `$\Omega / \square$' to avoid confusion with the raw four-terminal resistance.} The $R_\text{xx}(T)$ behavior shown in the inset of Fig.~\ref{fig:2}(a) is typical for a bulk-insulating (Bi$_\text{x}$Sb$_\text{1-x}$)$_2$Te$_3$ film \cite{Yang2015}. The $R_\text{xx}(V_{\rm G})$ behavior shown in the main panel of Fig.~\ref{fig:2}(a) shows a clear peak with the gate-tunabiliy of $R_\text{xx}$ by a factor of 2, which further confirms the bulk-insulation and indicates that the chemical potential can be gate-tuned across the Dirac point of the topological surface state. 

For each data point in the gating curve, we measured $R_\text{xx}$ and $R_\text{yx}$ as a function of the perpendicular magnetic field up to $9\,$T.  By employing the two-band analysis used in Refs. \cite{Ren2010,Ando2013,Yang2015}, we analysed the sheet carrier density $n_\text{2D}$ and  the mobility $\mu$ of the carriers contributing to the transport at each $V_\text{G}$ [see Fig.~S1 in the Supporting information (SI) for the Hall data] and the results are shown in Figs.~\ref{fig:2}(b) and \ref{fig:2}(c). We identified a single $n$-type band for $V_{\rm G} > -4\,$V, while both $n$- and $p$-type carriers contribute to the transport for $V_{\rm G} \leq -4\,$V. The mobility of the $n$-type carriers is of the order of 1000 cm$^2$/Vs which likely comes from the surface electrons in both top and bottom surfaces \cite{Yang2015}. At $V_{\rm G} \leq -4\,$V, the top surface is tuned into the $p$-type regime where holes from both surface (2D) and bulk (3D) contribute to the transport (because of the close proximity of the bulk valence-band edge to the Dirac point of the surface state \cite{Kong2011, Zhang2011}), while the bottom surface is still in the $n$-type regime and keep contributing high-mobility electrons. This behavior is entirely consistent with what was reported for top-gated (Bi$_\text{x}$Sb$_\text{1-x}$)$_2$Te$_3$ films grown on pristine sapphire substrates \cite{Yang2015}.

\begin{figure}
\centering
\includegraphics[width=.45\textwidth]{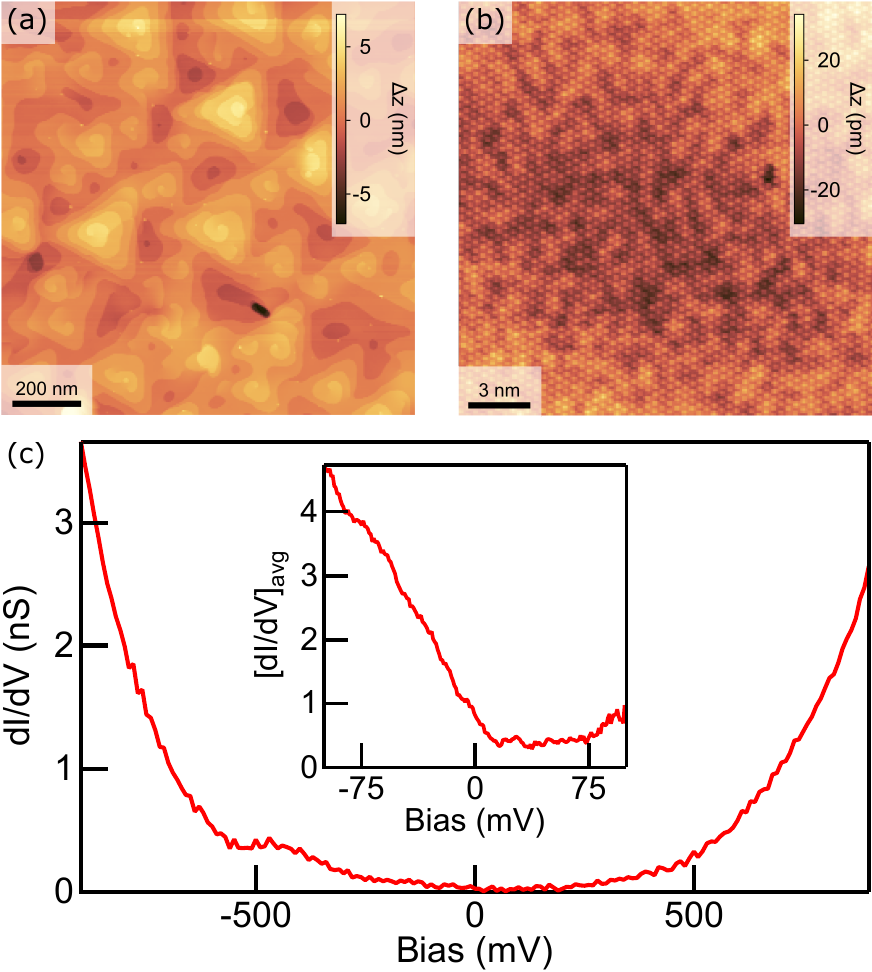}
\vspace*{0mm}\caption{\GL{STM/STS characterization of the SAE-grown film on substrate 2. (a) Topographic image showing triangular terraces with the step height of one quintuple layer ($\sim$1$\,$nm). (b) Atomic-resolution image showing the hexagonal lattice of the top Te-layer; the color contrast is due to the random
arrangement of Bi/Sb atoms in the subsurface layer \cite{Knispel2017}. (c) Example d$I$/d$V$ spectrum for a wide bias range. Inset shows the averaged spectrum $\left[dI/dV\right]_{\rm avg}$, for which individual d$I$/d$V$ spectra at various points are first normalized by the setpoint conductance $I_0/V_0$ (see Methods) and then averaged.}}
\label{fig:3}
\end{figure}

\GL{To obtain further evidence for bulk-insulation, we performed scanning tunneling microscopy and spectroscopy (STM/STS) on a SAE-grown thin film (see Methods for details). Figures \ref{fig:3}(a) and \ref{fig:3}(b) show topographic and atomic-resolution images of substrate 2, respectively, which give additional evidence for the good epitaxial growth. To gain access to the local density of states (DOS), the differential conductance d$I$/d$V$ was measured as a function of bias voltage on several locations on the SAE-grown thin film. A representative d$I$/d$V$ spectrum is shown in Fig.~\ref{fig:3}(c) for a wide bias range, and the inset shows an averaged spectrum close to zero bias. Apparently, the chemical potential (i.e. zero bias) lies close to the minimum of the DOS, which gives clear evidence for bulk insulation. }

While the (Bi$_\text{x}$Sb$_\text{1-x}$)$_2$Te$_3$ film grown in the Hall-bar structure analyzed above has shown bulk-insulation with the tunability of the chemical potential across the Dirac point, the 160-nm-wide nanowire structure grown on the very same substrate turned out to be different. Within the range of $V_{\rm G}$ that can be applied without breaking the dielectric, the Dirac point cannot be accessed and the $R_\text{xx}(V_{\rm G})$ indicated that the nanowire was always in the $p$-type regime [Fig.~\ref{fig:2}(d)]. This suggests that the MBE-growth condition depends on the size of the patterned features and the actual composition of (Bi$_\text{x}$Sb$_\text{1-x}$)$_2$Te$_3$ is different between the Hall-bar and nanowire structures, which is consistent with a past report on SAE \cite{Jalil2023}. The lower value of the sheet resistance in the nanowire [Fig.~\ref{fig:2}(d)] compared to that in the 10-$\mu$m-wide Hall-bar device on the same substrate [Fig.~\ref{fig:2}(d)] also supports this conclusion. Since this nanowire is bulk-conducting, the AB-like magnetoresistance oscillations characteristic of bulk-insulating TI nanowires \cite{Peng2010} are not recognizable in parallel magnetic fields [see inset of Fig.~\ref{fig:2}(d) and additional data in SI].

\begin{figure*}
\centering
\includegraphics[width=.85\textwidth]{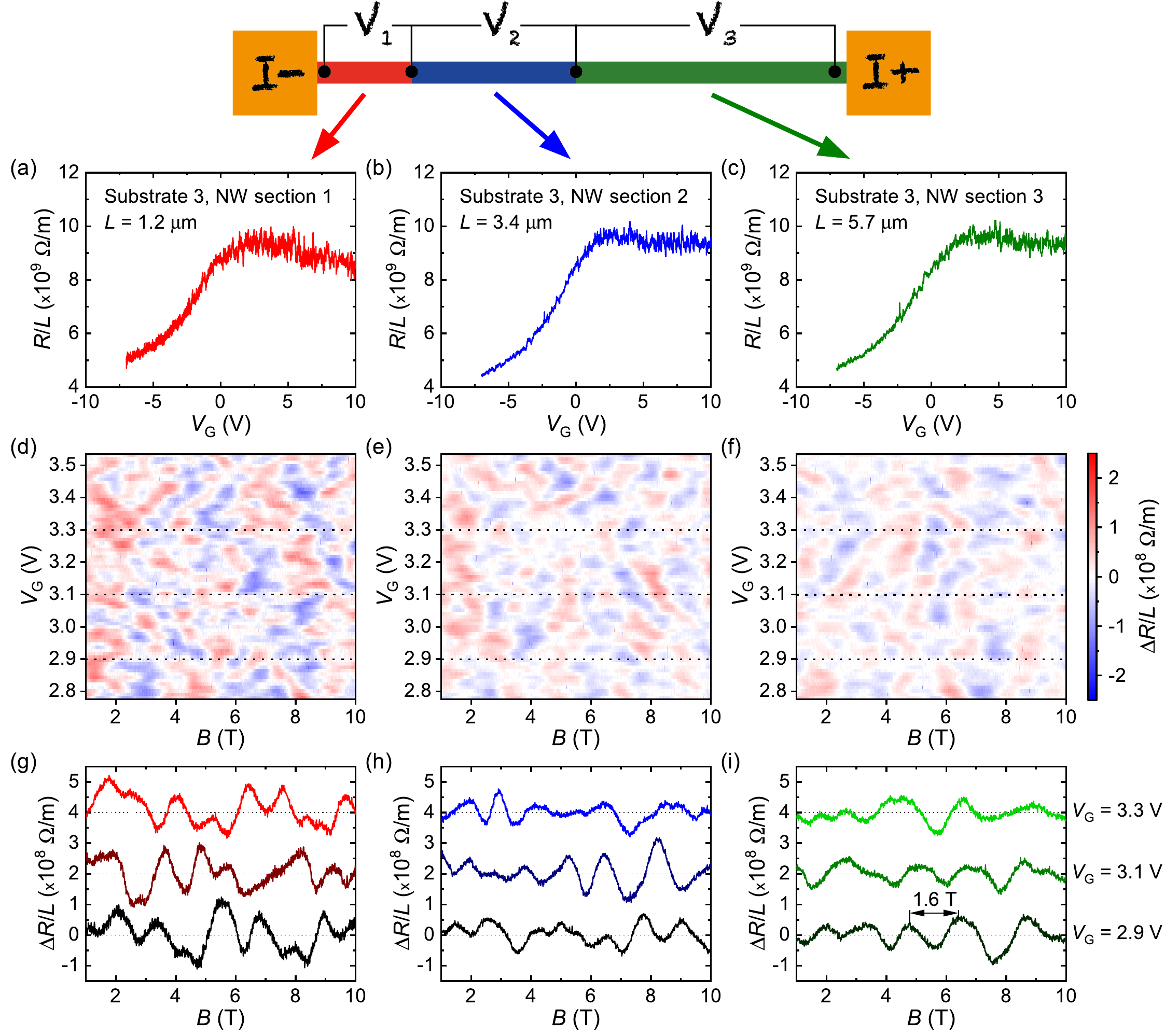}
\vspace*{0mm}\caption{Comparison of the transport properties of three different sections of the 120-nm-wide bulk-insulating (Bi$_\text{x}$Sb$_\text{1-x}$)$_2$Te$_3$ nanowire (NW) SAE-grown on substrate 3. (a-c) $V_\text{G}$-dependence of the resistance $R$ normalized by the section length $L$ measured at $260\,$mK in $0\,$T for (a) $L = 1.2\,\mu$m, (b) $L = 3.4\,\mu$m, and (c) $L = 5.7\,\mu$m. The schematic of the measurement set-up is shown in the top. (d-f) 2D maps of the magnetoresistance oscillations in parallel magnetic fields as functions of $B$ and $V_{\rm G}$ after subtracting a smooth background (see SI for details) for the three sections. (g-i) Background-subtracted magnetoresistance curves as a function of the parallel magnetic field, measured at $V_\text{G} = 2.9\,$V, $3.1\,$V, and $3.3\,$V for the three sections; the positions of these $V_\text{G}$ values are marked in (d-f) with black dashed lines.}
\label{fig:4}
\end{figure*}

To achieve bulk-insulation in SAE-grown nanowires, the MBE growth conditions were reoptimised for substrate 3, decreasing the beam-equivalent-pressure ratio of Sb/Bi from 2.1 to 1.8 (see Methods for details). For a long (Bi$_\text{x}$Sb$_\text{1-x}$)$_2$Te$_3$ nanowire grown on substrate 3, three sections with different lengths were investigated. Figures \ref{fig:4}(a-c) show the $V_{\rm G}$-dependence of the resistance $R$ normalized by the section length $L$ for the three sections 1, 2 and 3. All three section can be tuned to the Dirac point with $V_\text{G} \approx 2.4\,$V and their resistances scale linearly with the wire length as expected for nanowires in the diffusive regime, demonstrating the homogeneity of the grown nanowire. 

In this bulk-insulating nanowire, we observed the characteristic AB-like magnetoresistance oscillations \GL{with a period of $h/e$} in parallel magnetic fields stemming from the magnetic-flux-dependence of the Dirac subbands formed in TI nanowires due to the quantum size effect \cite{Peng2010, Xiu2011, Cho2015, Roessler2023}. The phase shift of these oscillations upon gating is commonly observed in bulk-insulating TI nanowires \cite{Cho2015, Kim2020, Roessler2023}, and it is also the case here: After subtracting a smooth background from the raw data to obtain $\Delta R$ (see SI for details), we find oscillations whose phase shifts with $V_{\rm G}$ [Figs.~\ref{fig:4}(d-f)]. Note that the oscillations and the phase shift have considerable irregularity, which points to strong disorder. This is understandable, because the bulk-insulation in (Bi$_\text{x}$Sb$_\text{1-x}$)$_2$Te$_3$ is achieved by compensation doping, which leaves charged donors and acceptors \cite{Skinner2012, Ando2013, Borgwardt2016, Knispel2017, Breunig2017, Nandi2018, Lippertz2022}.

Interestingly, the degree to which disorder affects the regularity of flux-dependent oscillations seems to correlate with the nanowire length. While the mapping of $\Delta R(B, V_\mathrm{G})$ shown in Fig.~\ref{fig:4}(d) for the shortest section is significantly disordered, a more regular pattern emerges upon increasing the wire length, as one can see in Figs.~\ref{fig:4}(e-f). One can also see this qualitative change in the individual curves of $\Delta R(B)$ for fixed $V_\mathrm{G}$ values [Figs.~\ref{fig:4}(g-i) \GL{and additional data in SI}]. This length dependence probably arises because the random resistance fluctuations due to disorder are averaged out in longer sections. In the longest 5.7-$\mu$m section where the AB-like oscillations were most regular, we identify the dominant periodicity of $\sim$1.6~T, which corresponds to a cross-sectional area of $2.6 \times 10^{-15}\,$m$^2$ \GL{for the magnetic flux of $h/e$}. This value is only slightly smaller than the geometrical cross-section of the wire, $2.9 \times 10^{-15}\,$m$^2$, which was determined via AFM and SEM observations of the nanowire thickness ($24\,$nm) and width ($120\,$nm), respectively. The flux-relevant cross-section that is slightly reduced from the geometrical value was previously attributed to the penetration depth of the surface-state wave function and a finite oxide shell on the surface \cite{Zhang2010, Rosenbach2020}.

\section{Conclusion}

\GL{One of the bottlenecks for the continued development of TI-based devices is the Coulomb disorder \cite{Beidenkopf2011, Chong2020, Okada2012, Fu2013, Lee2015, Pauly2015, Fu2016, Storz2016, Knispel2017}, which is unavoidable in material like (Bi$_\text{x}$Sb$_\text{1-x}$)$_2$Te$_3$ where the bulk-insulation is achieved by compensation \cite{Ando2013, Skinner2013}. The Coulomb disorder is particularly problematic for achieving topological quantum computing using Majorana zero modes \cite{Cook2011, Huang2021}. The solution of this bottleneck probably requires discoveries of new materials or new growth techniques. Another bottleneck is the generally  poor interface with other materials such as superconductors or ferromagnets \cite{Breunig2022}. The poor interface limits the performance of the devices that are based on the properties of the topological surface states. The SAE technique is useful for solving this bottleneck, because epitaxial growth of superconductors or ferromagnets {\it in-situ} on SAE-grown TI structures \cite{Schueffelgen2019} would allow for fabrication of functional devices having a high-quality epitaxial interface.}

In this work, we have shown that bulk-insulation can be achieved in TI structures that are SAE-grown on a patterned sapphire substrate with a Si$_3$N$_4$ growth mask. The high carrier mobility and the tunability of the chemical potential across the Dirac point demonstrate the suitability of this technique to engineer more complex devices. We found that the optimum growth conditions to achieve bulk-insulation changes with the feature size, but it was still possible to optimize the conditions to grow a bulk-insulating (Bi$_\text{x}$Sb$_\text{1-x}$)$_2$Te$_3$ nanowire with the width of only 120 nm. The AB-like magnetoresistance oscillations observed in such a nanowire confirm the quasi-1D transport through the quantum-confined topological surface state. Interestingly, our data suggest that the disorder effect, which causes random fluctuations of the wire resistance, tends to be averaged out in longer wires. The present result opens the possibility to fabricate TI-based devices to exploit various topological phenomena in a scalable manner.

\section{Methods}

\begin{flushleft}
{\bf MBE growth \& device fabrication:}\\
\end{flushleft}
\vspace{-3mm}
The SAE growth templates were created by depositing a 50-nm Si$_3$N$_4$ layer by hot-wire chemical vapour deposition (HW-CVD) on top of $1 \times 1\,$cm$^2$ sapphire (0001) substrates. Subsequently, the growth masks were defined using electron-beam lithography, and transferred into the Si$_3$N$_4$ layer by reactive-ion etching with CF$_4$. The resist was removed in acetone and isopropanol. Next, the template was exposed to a mild Ar and O$_2$ plasma to remove any remaining resist leftovers.

The growth templates were then loaded into the MBE chamber (MBE Komponenten, Octoplus 400) and annealed up to 900$^{\circ}$C in ultra-high vacuum to clean off any adsorbents. Good thermal contact between the substrate and sample holder was found to be crucial for successful SAE growth, and was achieved by spot-welding all four corners of the substrate to the Mo sample holder using Ta strips. For the MBE growth of (Bi$_\text{x}$Sb$_\text{1-x}$)$_2$Te$_3$, high-purity Bi, Sb, and Te source materials (99.9999\%, Thermo Fisher) were co-evaporated onto the substrate. For substrate 1, the beam-equivalent-pressures (BEP) for Bi, Sb, and Te were $2.20 \times 10^{-8}$, $4.60 \times 10^{-8}$, and $1.00 \times 10^{-6}\,$mBar respectively, which yielded a bulk-insulating Hall-bar, but $p$-type metallic nanowire. Te was supplied in excess, while Bi and Sb acted as the limiting species. The BEP ratio of Te/(Bi+Sb) was $\sim$15 and Sb/Bi was $\sim$2.1. \GL{For substrate 2 (STM/STS study), the BEP for Bi, Sb, and Te were $2.40 \times 10^{-8}$, $4.70 \times 10^{-8}$, and $1.00 \times 10^{-6}\,$mBar, respectively [Te/(Bi+Sb) $\approx$ 14 and Sb/Bi $\approx$ 2.0].} For substrate 3, the BEP for Bi, Sb, and Te were $2.70 \times 10^{-8}$, $4.90 \times 10^{-8}$, and $1.00 \times 10^{-6}\,$mBar [Te/(Bi+Sb) $\approx$ 13 and Sb/Bi $\approx$ 1.8], where the reduced Sb/Bi BEP ratio yielded bulk-insulating nanowires with widths in the range of $100\,$nm.

The substrate temperature was kept at 245$^{\circ}$C for the first $3\,$min of the MBE deposition, during which a continuous seed layer of (Bi$_\text{x}$Sb$_\text{1-x}$)$_2$Te$_3$ formed on the exposed sapphire (0001) surface. Almost no nucleation of (Bi$_\text{x}$Sb$_\text{1-x}$)$_2$Te$_3$ crystallites occurred on the Si$_3$N$_4$ growth mask as the sticking coefficients for Bi, Sb, and Te at 245$^{\circ}$C were too low for this surface. Next, the substrate temperature was slowly increased to 285$^{\circ}$C in 13 min (i.e. 3$^{\circ}$C/min), while the deposition continued. At 285$^{\circ}$C the substrate was kept for an additional $20\,$min in the fluxes of Bi, Sb, and Te, after which the substrate was left to cool down naturally to room temperature in zero flux. The higher growth temperature of 285$^{\circ}$C lead to a better morphology of the (Bi$_\text{x}$Sb$_\text{1-x}$)$_2$Te$_3$ film than would be obtained from growing with a constant temperature profile of 245$^{\circ}$C. Note, however, that a lower substrate temperature of 245$^{\circ}$C at the start was needed for the formation of a continuous seed layer of (Bi$_\text{x}$Sb$_\text{1-x}$)$_2$Te$_3$ which does not take place at 285$^{\circ}$C on sapphire (0001) (and Si$_3$N$_4$) for the Bi, Sb, and Te fluxes used in our MBE system.

Within a few minutes of taking the films out of the MBE chamber, substrate 1 and 3 were capped with a 40-nm thick Al$_2$O$_3$ capping layer grown by atomic layer deposition (ALD) at 80$^{\circ}$C using Ultratec Savannah S200 to avoid degradation in air. Next, the (Bi$_\text{x}$Sb$_\text{1-x}$)$_2$Te$_3$ contact arms of the Hall-bar and nanowire devices were metallized by sputter-depositing 5 nm of Pt and 45 nm of Au. For this process, optical lithography was used and the Al$_2$O$_3$ capping layer was selectively removed in the contact areas by an aluminum etchant (Transene Type-D) heated to 50$^{\circ}$C. Lastly, a top-gate electrode was defined on top of the 40-nm thick Al$_2$O$_3$ layer by electron-beam lithography and sputter-depositing 5 nm Pt + 45 nm Au. \GL{Substrate 2, on the other hand, was capped by $\sim 10\,$nm of Te at room temperature inside the MBE chamber to protect the films during the ex-situ transfer to the STM system.} The dimensions of the devices studied in this work are summarized in the supplement.

\begin{flushleft}
{\bf \GL{Magneto-transport} measurements:}\\
\end{flushleft}
\vspace{-3mm}
The characterization of the selectively-grown Hall-bar device (substrate 1) was carried out in a Quantum Design physical property measurement system (PPMS), which can be operated in a temperature range from $2\,$K to $350\,$K and in magnetic fields up to $9\,$T. The measurements of the nanowire device (substrate 1) were performed in a dry dilution refrigerator (Oxford Instruments TRITON 200, base temperature $\sim$40mK) equipped with a 8-T superconducting magnet. The measurements of the magnetoresistance oscillations in the nanowire on substrate 3 were performed in a $^3$He cryostat (Oxford Instrument Heliox) at a base temperature of $250\,$mK in magnetic fields up to $10\,$T. The nanowire (Hall-bar) devices were measured using standard lock-in techniques in a four-terminal configuration with an ac excitation current of $10\,$nA ($1\,\mu$A) and a lock-in frequency of $13.37\,$Hz.

\GL{
\begin{flushleft}
{\bf Scanning tunneling microscopy/spectroscopy:}\\
\end{flushleft}
\vspace{-3mm}
Inside the STM preparation chamber, the sample (substrate 2) was first outgassed at $400\,$K for $30\,$min to desorb the water molecules, and subsequently heated up to $540\,$K in $10\,$min to desorb the Te capping layer. After $5\,$min at $540\,$K, the sample was left to cool down to room temperature. Next, the sample was transferred in-situ into the STM main chamber (Unisoku USM1300) and cooled down to $1.7\,$K in UHV. The topographic (atomic-resolution) map was performed in the constant-current mode with the setpoint values $I_0 = 50\,$pA ($0.5\,$nA) and $V_0 = 2\,$V ($-0.3\,$V). The spectroscopy data are obtained by first stabilizing for a given setpoint condition [e.g., $I_0 = 0.5\,$nA and $V_0 = 0.9\,$V for the d$I$/d$V$ spectrum in the main panel of Fig.~\ref{fig:3}(c)] and then disabling the feedback loop. The setpoint differs at different positions. 
Since our SAE technique relies on the electrically insulating Si$_3$N$_4$ layer as the growth mask, the STM/STS characterization could only be performed on an SAE-grown film away from the Si$_3$N$_4$ edge to avoid crashing the STM tip.
}

\begin{flushleft}
{\bf Data and materials availability:}\\
\end{flushleft}
\vspace{-3mm}
The data used in the generation of main and supplementary figures are available in Zenodo with the identifier \href{https://doi.org/10.5281/zenodo.12544807}{10.5281/zenodo.12544807}.

\GL{
\section{Supporting Information}
Table: summary of the dimensions of the devices studied in this work,
Fig.~S1: Hall data for the two-band fit of substrate 1, 
Fig.~S2: magnetic-field dependence of the metallic nanowires on substrate 1, 
Fig.~S3: details of the background subtraction for substrate 3, and
Fig.~S4: additional magnetoresistance data for the bulk-insulating nanowire.
}

\section{Acknowledgements}
We would like to thank Ella Nikodem, Cornelius Dietrich, and Sandra Omoragbon for the technical support in creating the SAE growth templates. This work has received funding from the Deutsche Forschungsgemeinschaft (DFG, German Research Foundation) under CRC 1238-277146847 (subprojects A04 and B01) and also from the DFG under Germany’s Excellence Strategy -- Cluster of Excellence Matter and Light for Quantum Computing (ML4Q) EXC 2004/1-390534769, as well as from the DFG Project No.~398945897.  

\providecommand{\noopsort}[1]{}\providecommand{\singleletter}[1]{#1}%
\providecommand{\latin}[1]{#1}
\makeatletter
\providecommand{\doi}
  {\begingroup\let\do\@makeother\dospecials
  \catcode`\{=1 \catcode`\}=2 \doi@aux}
\providecommand{\doi@aux}[1]{\endgroup\texttt{#1}}
\makeatother
\providecommand*\mcitethebibliography{\thebibliography}
\csname @ifundefined\endcsname{endmcitethebibliography}
  {\let\endmcitethebibliography\endthebibliography}{}

\end{document}



\title{Supporting Information \\Selective-Area Epitaxy of Bulk-Insulating (Bi$_\text{x}$Sb$_\text{1-x}$)$_2$Te$_3$ Films and Nanowires by Molecular Beam Epitaxy}

\author{Gertjan Lippertz}
\affiliation{Physics Institute II, University of Cologne, D-50937 K{\"o}ln, Germany}

\author{Oliver Breunig}
\affiliation{Physics Institute II, University of Cologne, D-50937 K{\"o}ln, Germany}

\author{Rafael Fister}
\affiliation{Physics Institute II, University of Cologne, D-50937 K{\"o}ln, Germany}

\author{Anjana Uday}
\affiliation{Physics Institute II, University of Cologne, D-50937 K{\"o}ln, Germany}

\author{Andrea Bliesener}
\affiliation{Physics Institute II, University of Cologne, D-50937 K{\"o}ln, Germany}

\author{\GL{Jens Brede}}
\affiliation{Physics Institute II, University of Cologne, D-50937 K{\"o}ln, Germany}

\author{Alexey Taskin}
\affiliation{Physics Institute II, University of Cologne, D-50937 K{\"o}ln, Germany}

\author{Yoichi Ando} \email{ando@ph2.uni-koeln.de}
\affiliation{Physics Institute II, University of Cologne, D-50937 K{\"o}ln, Germany}

\maketitle

\section{\label{sec:samples}Summary of Devices}

\noindent The following SAE-grown (Bi$_\text{x}$Sb$_\text{1-x}$)$_2$Te$_3$ structures are reported in this paper:

\begin{table}[h]
    \centering
    \begin{tabular}{m{3.5cm}m{2cm}m{2cm}m{2cm}m{2cm}m{3cm}}
        \toprule
        Device & Substrate & Length & Width & Thickness & Expected AB-period \\
        \midrule
        Hall-bar & 1 & $64\,\mu$m & $10\,\mu$m & $20\,$nm & -- \\
        Nanowire & 1 & $8.0\,\mu$m & $160\,$nm & $20\,$nm & $1.3\,$T \\
        \GL{STM, Large window} & 2 & $315\,\mu$m & $400\,\mu$m & $28\,$nm & -- \\
        Nanowire, Section 1 & 3 & $1.2\,\mu$m & $120\,$nm & $24\,$nm & $1.5\,$T \\
        Nanowire, Section 2 & 3 & $3.4\,\mu$m & $120\,$nm & $24\,$nm & $1.5\,$T \\
        Nanowire, Section 3 & 3 & $5.7\,\mu$m & $120\,$nm & $24\,$nm & $1.5\,$T \\
        \bottomrule
    \end{tabular}
\end{table}


\section{\label{sec:hall-bar}Two-band analysis of the Hall data}

\begin{figure*}[t]
\centering
\includegraphics[width=.85\textwidth]{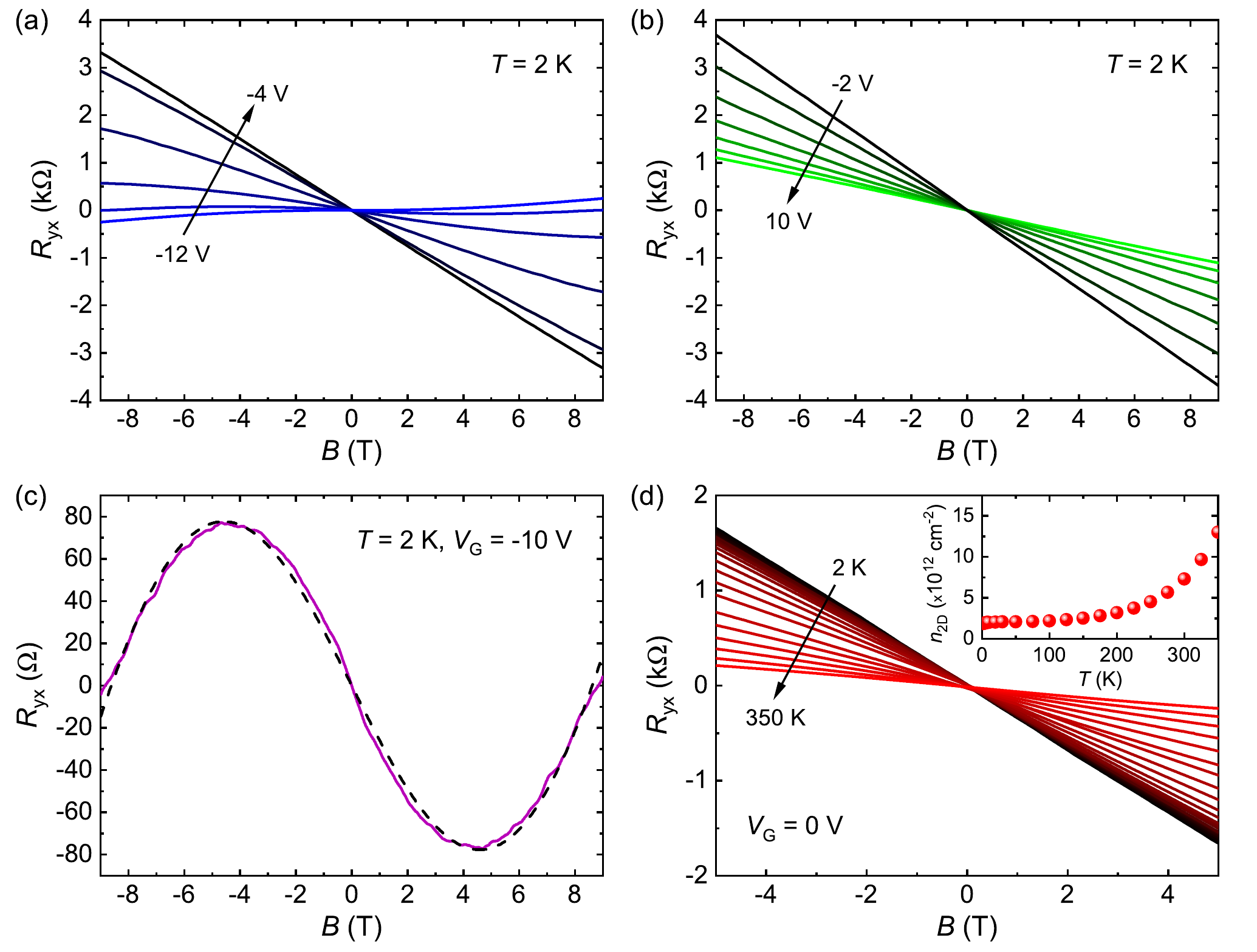}
\vspace*{0mm}\caption{(a,b) $R_\text{yx}(B)$ curves for the Hall-bar device on substrate 1 measured at 2 K for $V_{\rm G}$ from $-12$ to $-4$ V (a) and from $-2$ to $-10$ V (b); the $V_{\rm G}$ value for each curve corresponds to the data points in Fig. 2 of the main text. (c) Two-band-model fit to the $V_{\rm G} = -10$ V data using Eq. S1. (d) $R_\text{yx}(B)$ curves at $V_{\rm G} = 0$ V for temperatures from 2 to 350 K; inset shows the extracted $T$-dependence of $n_{\rm 2D}$. All $R_\text{yx}$ curves were antisymmetrized in $B$.}
\label{fig:S1}
\end{figure*}

In Fig.~2 of the main text, the $V_{\rm G}$-dependencies of the sheet carrier density $n_{\rm 2D}$ and mobility $\mu$ of the $n$- and $p$-type channels identified in the Hall-bar device are shown. To extract these values, the magnetic-field dependence of the Hall resistance $R_\text{yx}$ shown in Figs.~\ref{fig:S1}(a-b) were fitted using the two-band model \cite{Ren2010, Ando2013, Yang2015} 
\begin{equation}
    R_{yx} t = \frac{R_1 \rho_2^2 + R_2 \rho_1^2 + R_1 R_2 (R_1 + R_2) B^2}{(\rho_1 + \rho_2)^2 + (R_1 + R_2)^2 B^2} B,
        \label{eq:Ryx}
\end{equation}
where $R_i$ and $\rho_i$ are the Hall coefficient and the resistivity of the $i$-th channel, respectively, and $t$ is the thickness. Additionally, we used the measured sheet resistance $R_{xx}$ at $B$ = 0 to constrain the fitting parameters $\rho_1$ and $\rho_2$ with the relation
\begin{equation}
    R_{xx}(B=0) = \frac{\rho_1 \rho_2}{t(\rho_1 + \rho_2)}.
        \label{eq:Rxx}
\end{equation}
The values of $n_{\rm 2D}$ and $\mu$ for each channel can be calculated from $ n_{\rm 2D}^i = t/(eR_i)$ and $\mu_i  = t/(e n_{\rm 2D}^i \rho_i)$. 
Figure \ref{fig:S1}(c) shows the $R_\text{yx}(B)$ behavior (purple line) at $2\,$K for $V_\text{G} = -10\,$V, where the two-band nature of the transport was most apparent. The fit (dashed line) reproduces the data reasonably well. The $R_\text{xx}(B=0)$ values used for constraining the fits are shown in Fig.~2(a) of the main text. 

Figure \ref{fig:S1}(d) shows the $R_\text{yx}(B)$ behavior at $V_{\rm G}$ = 0 taken for 2 -- 350 K. At this $V_{\rm G}$ value, the $R_\text{yx}(B)$ behavior is essentially linear and single-band analysis yields the $T$-dependence of $n_{\rm 2D}$ shown in the inset.

\section{\label{sec:mNW}Additional magnetoresistance data for the bulk-conducting nanowire}

\begin{figure}[h]
\centering
\includegraphics[width=.85\textwidth]{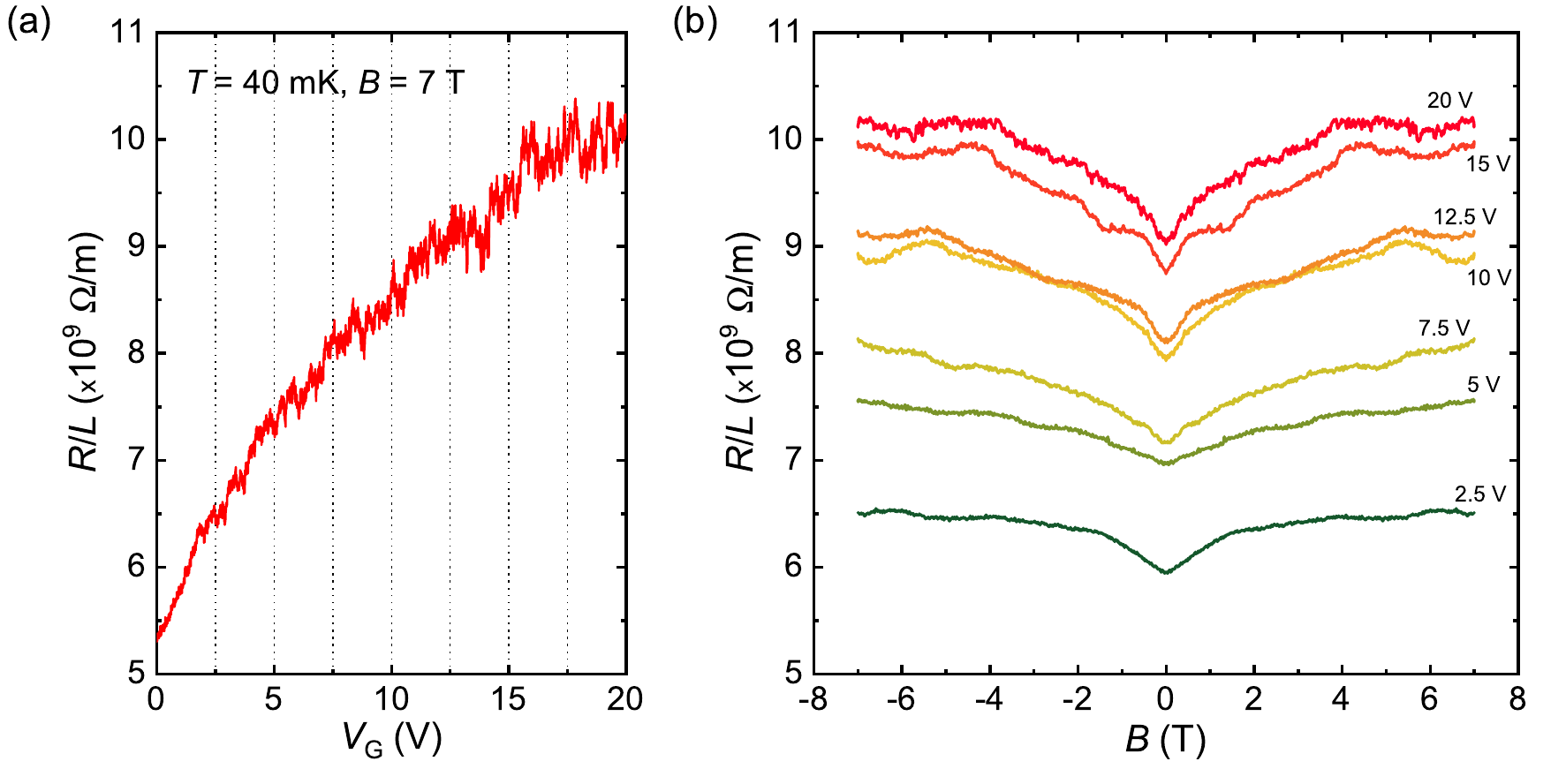}
\vspace*{0mm}\caption{Additional data for the bulk-conducting nanowire device on substrate 1. (a) $V_{\rm G}$ dependence of the resistance $R$ normalized by the length $L$ measured at $40\,$mK in the parallel magnetic field of $7\,$T. (b) Magnetic-field dependence of $R/L$ measured in parallel magnetic fields at various $V_\text{G}$ values at $40\,$mK. The $R/L$ curves were antisymmetrized in $B$. No regular oscillations can be identified.}
\label{fig:S2}
\end{figure}

In Fig.~2(d) of the main text, the magnetic-field dependence of the sheet resistance $R_\text{xx}$ at $V_\text{G} = 20\,$V is shown for the 8.0-$\mu$m-long nanowire device on substrate 1. Even though the Dirac point was not reached, electrostatic gating experiments at higher $V_\text{G}$ were not carried out, because the ALD-deposited Al$_2$O$_3$ layer is at risk of dielectric breakdown when exposed to electric fields higher than $\sim 0.5\,$V/nm. Additional magnetoresistance data at lower $V_\text{G}$ values are shown in Fig.~\ref{fig:S2}. One can see some features showing up as $V_\text{G}$ is increased (i.e. as the chemical potential is moved towards the Dirac point), but they are irregular and no periodic oscillations could be identified.

\clearpage

\section{\label{sec:background}Background subtraction}

\begin{figure}[h]
\centering
\includegraphics[width=0.6\textwidth]{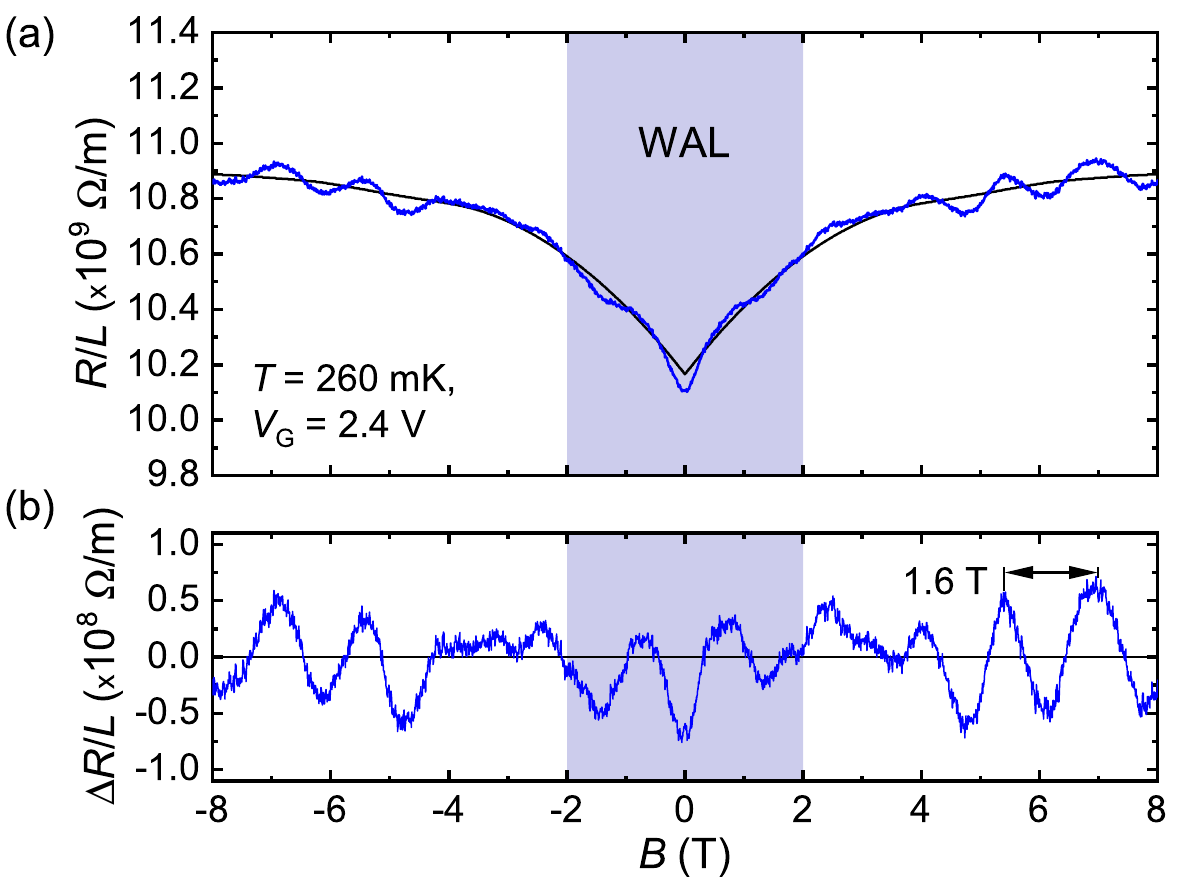}
\vspace*{0mm}\caption{Example of the background-subtraction procedure. (a) $R(B)$ data at the Dirac point ($V_{\rm G}$ = 2.4 V) for the $L = 3.4\,\mu$m section of the nanowire on substrate 3 (blue) normalized by the length $L$, measured at 260 mK. The background determined by LOESS-filtering is shown in black. The blue-shaded range is strongly affected by the weak antilocalization (WAL) effect. (b) Background-subtracted data in which the oscillation period of $\sim$1.6~T can be identified at high magnetic fields.}
\label{fig:S3}
\end{figure}

When analyzing Aharonov-Bohm (AB)-like oscillations in TI nanowires, because the weak antilocalization (WAL) effect \cite{Ando2013} and positive magnetoresistance are superimposed on these oscillations, it is customary to subtract a smooth background from the magnetoresistance data \cite{Rosenbach2020, Kim2020, Rosenbach2022, Roessler2023} to make the oscillations more apparent. In this work, the smooth background was determined by applying a LOESS (locally estimated scatterplot smoothing) filter to the $R(B)$ data in the ranges from $-10$ T to 0 T and from 0 T to 10 T separately, with an averaging window of 1497 data points (which corresponds to a span of about $7.5\,$T). Figure \ref{fig:S3} shows an example of the raw data, LOESS-generated background, and background-subtracted data $\Delta R$ divided by the length $L$. The data in the range from $-2$ to $2\,$T [shaded area in Fig.~\ref{fig:S3}] are strongly affected by the WAL, but the higher-field portion of the data show oscillation with a well-defined period of $\sim$1.6~T, which matches well to the cross-sectional area of the nanowire as discussed in the main text. 

\GL{
\section{\label{sec:BI-NW}Additional magnetoresistance data for bulk-insulating nanowire}
In Fig.~4 of the main text, pronounced AB-like magnetoresistance oscillations were shown for the three nanowire sections on substrate 3 in the gate-voltage range from $2.8\,$V to $3.5\,$V. Figure S4 shows additional data for slightly higher and lower gate voltages. The AB-like oscillations become more irregular and weaker when the chemical potential is moved away from the Dirac point. 
}

\newpage

\begin{figure}
\centering
\includegraphics[width=\textwidth]{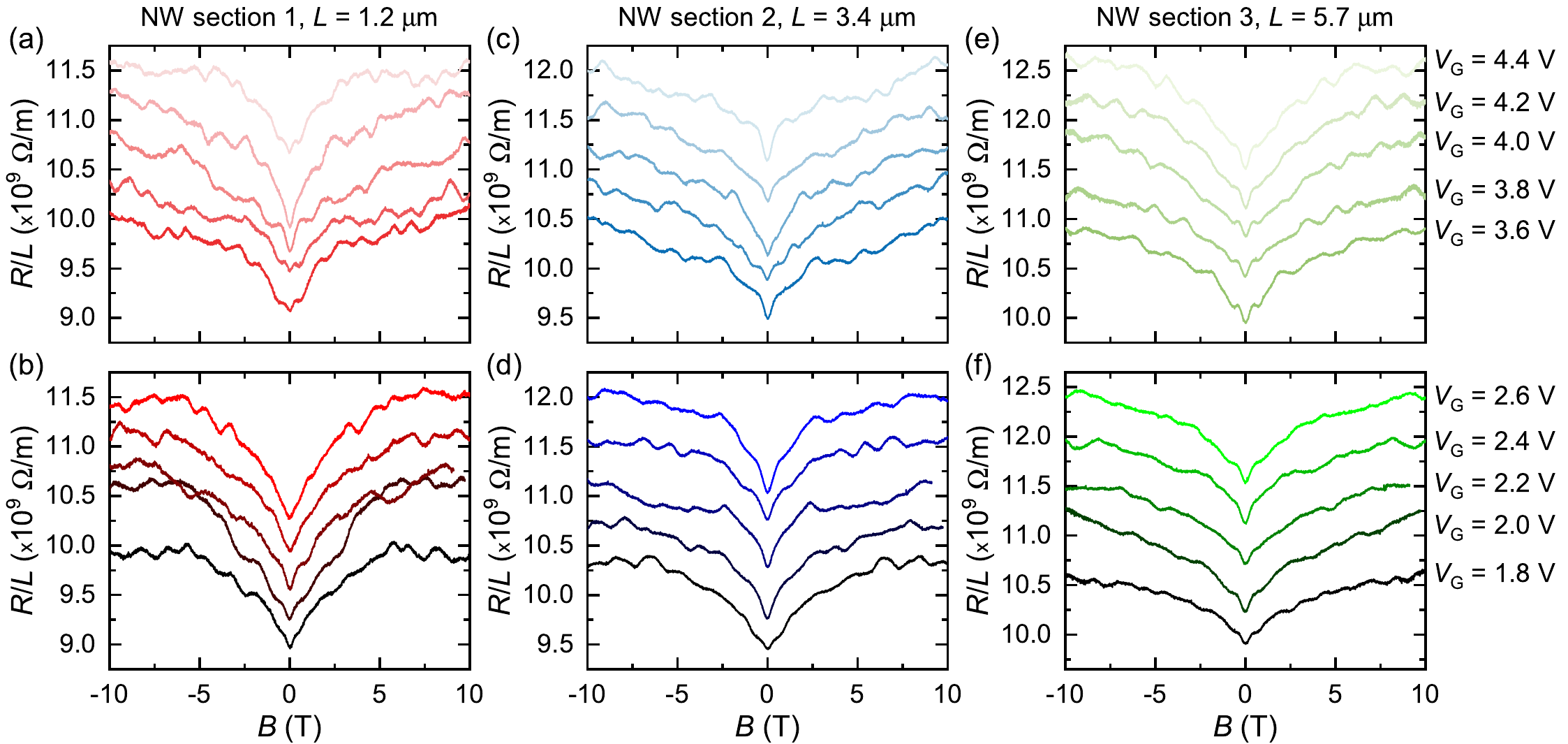}
\GL{\vspace*{0mm}\caption{Additional data for the three bulk-insulating nanowire (NW) sections on substrate 3. (a-f) Magnetic-field dependence of $R/L$ at $260\,$mK measured in parallel magnetic fields at various $V_\text{G}$ values [indicated on the right of (e) and (f)], for $L = 1.2\,\mu$m (a-b), $L = 3.4\,\mu$m (c-d), and $L = 5.7\,\mu$m (e-f). The $R/L$ curves for $V_{\rm G}$ = 1.8 V and 3.6 V were plotted without shifting, but the other curves were shifted vertically for clarity. Magnetoresistance oscillations are visible for several $V_\text{G}$-values in the raw data, even without subtracting a smooth background. The $R/L$ data is essentially symmetric in the magnetic field $B$, as expected.}}
\label{fig:S4}
\end{figure}

\providecommand{\noopsort}[1]{}\providecommand{\singleletter}[1]{#1}%
\providecommand{\latin}[1]{#1}
\makeatletter
\providecommand{\doi}
  {\begingroup\let\do\@makeother\dospecials
  \catcode`\{=1 \catcode`\}=2 \doi@aux}
\providecommand{\doi@aux}[1]{\endgroup\texttt{#1}}
\makeatother
\providecommand*\mcitethebibliography{\thebibliography}
\csname @ifundefined\endcsname{endmcitethebibliography}
  {\let\endmcitethebibliography\endthebibliography}{}